\documentclass[aps,prb,amsmath,amssymb,preprint,showpacs,longbibliography]{revtex4-1}

\usepackage{graphicx}
\usepackage{color}

\newlength{\largeur}
\newlength{\doublelargeur}

\newcommand{\degree}{\ensuremath{^\circ}}
\newcommand{\TM}{\ensuremath{T_\textrm{M}}}
\newcommand{\TC}{\ensuremath{T_\textrm{C}}}
\newcommand{\TA}{\ensuremath{T_\textrm{A}}}

\newcommand{\PSI}{\affiliation{Swiss Light Source, Paul Scherrer Institut, CH-5232 PSI-Villigen, Switzerland}}
\newcommand{\RU}{\affiliation{Radboud University Nijmegen, Institute for Molecules and Materials, 6525 AJ Nijmegen, The Netherlands}}
\newcommand{\CST}{\affiliation{College of Science and Technology, Nihon University, 24-1 Narashinodai 7-chome, Funabashi-shi, Chiba 274-8501, Japan}}
\newcommand{\aCST}{\altaffiliation{Present address: College of Science and Technology, Nihon University, 24-1 Narashinodai 7-chome, Funabashi-shi, Chiba 274-8501, Japan}}
\newcommand{\BESSY}{\affiliation{Helmholtz-Zentrum Berlin f\"{u}r Materialien und Energie GmbH, Albert-Einstein-Strasse 15, 12489 Berlin, Germany}}
\newcommand{\aETH}{\altaffiliation{Present address: Institute for Quantum Electronics, Physics Department, ETH Zurich, CH-8093 Zurich, Switzerland}}

\begin{document}

\title{All-optical magnetization switching in ferrimagnetic alloys: deterministic vs thermally activated dynamics}

\author{L. Le Guyader}
\email{email: loic.le\_guyader@helmholtz-berlin.de}
\BESSY{}
\PSI{}

\author{S. El Moussaoui}
\aCST{}
\PSI{}

\author{M. Buzzi}
\PSI{}

\author{M. Savoini}
\aETH
\RU{}

\author{A. Tsukamoto}
\author{A. Itoh}
\CST{}

\author{A. Kirilyuk}
\author{Th. Rasing}
\RU{}

\author{F. Nolting}
\PSI{}

\author{A. V. Kimel}
\RU{}


\date{\today}

\begin{abstract}
Using photo-emission electron microscopy with X-ray magnetic circular dichroism
as a contrast mechanism, new insights into the all-optical magnetization
switching (AOS) phenomenon in GdFe based rare-earth transition metal
ferrimagnetic alloys are provided. From a sequence of static images taken after
single linearly polarized laser pulse excitation, the repeatability of AOS
can be measured with a correlation coefficient.  It is found that
low coercivity enables thermally activated domain wall motion, limiting in
turn the repeatability of the switching. Time-resolved measurement of the
magnetization dynamics reveal that while AOS occurs below and above the
magnetization compensation temperature \TM{}, it is not observed in GdFe
samples where \TM{} is absent.  Finally, AOS is experimentally demonstrated
against an applied magnetic field of up to 180~mT.
\end{abstract}

%
\maketitle 

\section{Introduction}

Controlling magnetism on the ultrashort time scale of sub-100~ps has become an
important research subject, not only for the potential applications in novel
high density and high speed magnetic recording technologies but also for the
unique opportunity to investigate magnetism on the fundamental time scales of
the interactions between electrons, spins and lattice.\cite{Stoehr2006} The
demonstration in 1996\cite{Beaurepaire1996} of a rather unexpected ultrafast
sub-ps demagnetization in a thin Ni film upon femtosecond laser excitation
inspired a large number of following studies.\cite{Kirilyuk2010} Of particular
importance was the surprising demonstration of a deterministic magnetization
reversal by the sole action of a single 40~fs laser pulse in GdFeCo rare
earth-transition metal (RE-TM) alloys.\cite{Stanciu2007} The microscopic
mechanism responsible for this phenomenon, now referred to as all-optical
switching (AOS), remains debated.  Element selective studies of the ultrafast
demagnetization in GdFeCo alloys led to the interpretation that AOS is driven
by the heating from the laser pulse and is therefore independent of the laser
polarization and largely insensitive to any applied magnetic
field.\cite{Radu2011, Ostler2012} The helicity dependent AOS reported earlier
could then be understood in terms of a differential light absorption induced by
the magnetic circular dichroism in the magnetic alloy.\cite{Khorsand2012}
Finally, as these RE-TM alloys usually display chemical inhomogeneities, the
role of super-diffusive spin currents is also being discussed.\cite{Graves2013}

While early studies concentrated on GdFeCo alloys, magnetization switching by
laser pulses has now been reported in a growing range of systems, namely other
RE-TM alloys~\cite{Alebrand2012} and multi-layers,\cite{Mangin2014} RE-free
synthetic ferrimagnets~\cite{Evans2014, Mangin2014} and granular
ferromagnets.\cite{Lambert2014} These recent developments are raising a number
of crucial questions for the understanding of the AOS phenomenon and its
transfer to real world applications. Among these, the exact role played by the
magnetization compensation temperature \TM{} at which the magnetization of the
two sub-lattices cancel each other remains a puzzle. On the one hand, strong changes
in the magnetization dynamics upon crossing \TM{} have been
observed~\cite{Vahaplar2009, Vahaplar2012, Medapalli2012} and AOS seems to
occur preferably for alloys displaying a \TM{} which can be reached through
laser excitation.\cite{Alebrand2012, Alebrand2012b, Mangin2014} On the other
hand, atomistic simulations of the spin dynamics as well as experiments have
shown that AOS is feasible below and above \TM{}.\cite{Ostler2012} In addition,
helicity dependent magnetization switching in granular ferromagnets where no
\TM{} exists has been reported.\cite{Lambert2014} Finally, in view of potential
applications, it is crucial to be able to characterize to which extent AOS
is a deterministic process.

In this article, we investigate all-optical magnetization
switching in GdFe based alloys
using photo-emission electron microscopy (PEEM) with X-ray magnetic circular
dichroism (XMCD) as a contrast mechanism, allowing imaging of the
magnetic domain configuration with a spatial resolution of approximately
100~nm. Single linearly polarized laser pulses were used to excite a
multi-domain configuration at temperatures below and above the magnetization
compensation temperature \TM{} of the alloys.
Introducing the Pearson product-moment correlation coefficient on
series of XMCD images allows us to report a nearly purely
deterministic AOS in both cases. Extrinsic pulse to pulse laser pointing
stability and intrinsic finite domain sizes and thermally activated domain wall
motion are found to be the main limiting factors for a purely deterministic AOS.
Using time-resolved XMCD PEEM imaging of the magnetization dynamics
upon femtosecond laser excitation with 70~ps time resolution and approximately
200~nm spatial resolution, it is found that AOS can even be achieved against a
180~mT applied magnetic field. Finally,
strong reduction of the switching window above \TM{} is observed and
is partly related with the proximity of the Curie temperature \TC{} of
the sample.

\section{Methods}
\subsection{Time-resolved XMCD PEEM}

In order to resolve the magnetic domain configuration and its dynamics upon
AOS, the Elmitec photoemission electron microscope (PEEM) at the
Surface/Interface: Microscopy (SIM) beamline\cite{Flechsig2010} at the Swiss
Light Source (SLS) was used.  Employing the X-ray magnetic circular dichroism
(XMCD) effect at the Fe L$_3$ edge at 708 eV, a quantitative determination of
the Fe spin orientation with a 100~nm spatial resolution is
possible.~\cite{Nolting2010} From two images recorded with opposite X-ray
helicity, an asymmetry image is computed which contains only normalized
magnetic contrast information. Such image typically shows white or black regions
corresponding to magnetic domains with magnetizations of opposite directions
with respect to the X-ray propagation vector.\cite{Scholl2002} Time-resolved
measurements of the sample magnetization were performed by taking advantage of
the pulsed nature of the X-rays produced by the SLS via the gating of the
detection in synchronization to an isolated X-ray pulse. This scheme,
presented in detail in
Ref.~\onlinecite{LeGuyader2012}, allows stroboscopic pump-probe imaging of the
sample with a time resolution determined by the 70~ps Full Width at Half
Maximum (FWHM) temporal X-ray pulse length. At this time scale, both TM and RE
magnetizations are in equilibrium such that measuring the Fe sub-lattice is
sufficient to characterize the sample magnetization orientation.  The pump
laser pulses were produced by an XL-500 oscillator from Femtolasers Produktions
GmbH which are characterized by a wavelength of $\lambda$ = 800~nm, a pulse
duration of $\tau$ = 50~fs with an energy of 500~nJ per pulse at a 5.2 MHz
repetition rate. This repetition rate is then reduced by a Pockels cell in
combination with a crossed polarizer to match the 1.04 MHz repetition rate of
the isolated X-ray probe pulses. The linearly p-polarized laser pump pulses
were focused on the sample at a grazing incidence of 16\degree{} to a spot size
of about 30$\times$100~$\mu$m$^2$ FWHM.  The time overlap ($t = 0$) between the
laser and the X-ray pulse is unambiguously determined to better than $\pm$15~ps
by the sudden space charging~\cite{Mikkelsen2009, Buckanie2009} which is
induced by the laser pump pulse which reduces significantly the amount of
photo-emitted electrons collected by the microscope. Finally, the sample could
be cooled down with a flow of liquid nitrogen and the temperature measured with
a thermocouple attached to the sample holder.

\subsection{Samples}

The samples are grown on Si substrates to achieve fast cooling time during MHz
repetition rate experiments~\cite{Hassdenteufel2014} and are capped with a
3~nm Si$_3$N$_4$ layer to prevent oxidation. Three different samples have been
used for this study. The first sample of composition Si\slash AlTi(10~nm)\slash
Si$_3$N$_4$(5~nm)\slash Gd$_{25}$Fe$_{65.6}$Co$_{9.4}$(20~nm)\slash
Si$_3$N$_4$(3~nm) has a \TM{} of 260~K. The two other
samples are GdFe alloys of composition Si\slash Si$_3$N$_4$(5~nm)\slash
Gd$_{20}$Fe$_{80}$(30~nm)\slash Si$_3$N$_4$(3~nm) with a \TM{} below 10~K and
Si\slash Si$_3$N$_4$(5~nm)\slash Gd$_{24}$Fe$_{76}$(30~nm)\slash
Si$_3$N$_4$(3~nm) with a \TM{} above 500~K. In the rest of the paper, each
sample is referred to by a reduced notation consisting of the Gd content like
for example Gd25FeCo or Gd20Fe.

\section{Results}
\subsection{Single laser pulse excitation}

\begin{figure}[ht]
\includegraphics[width=\doublelargeur]{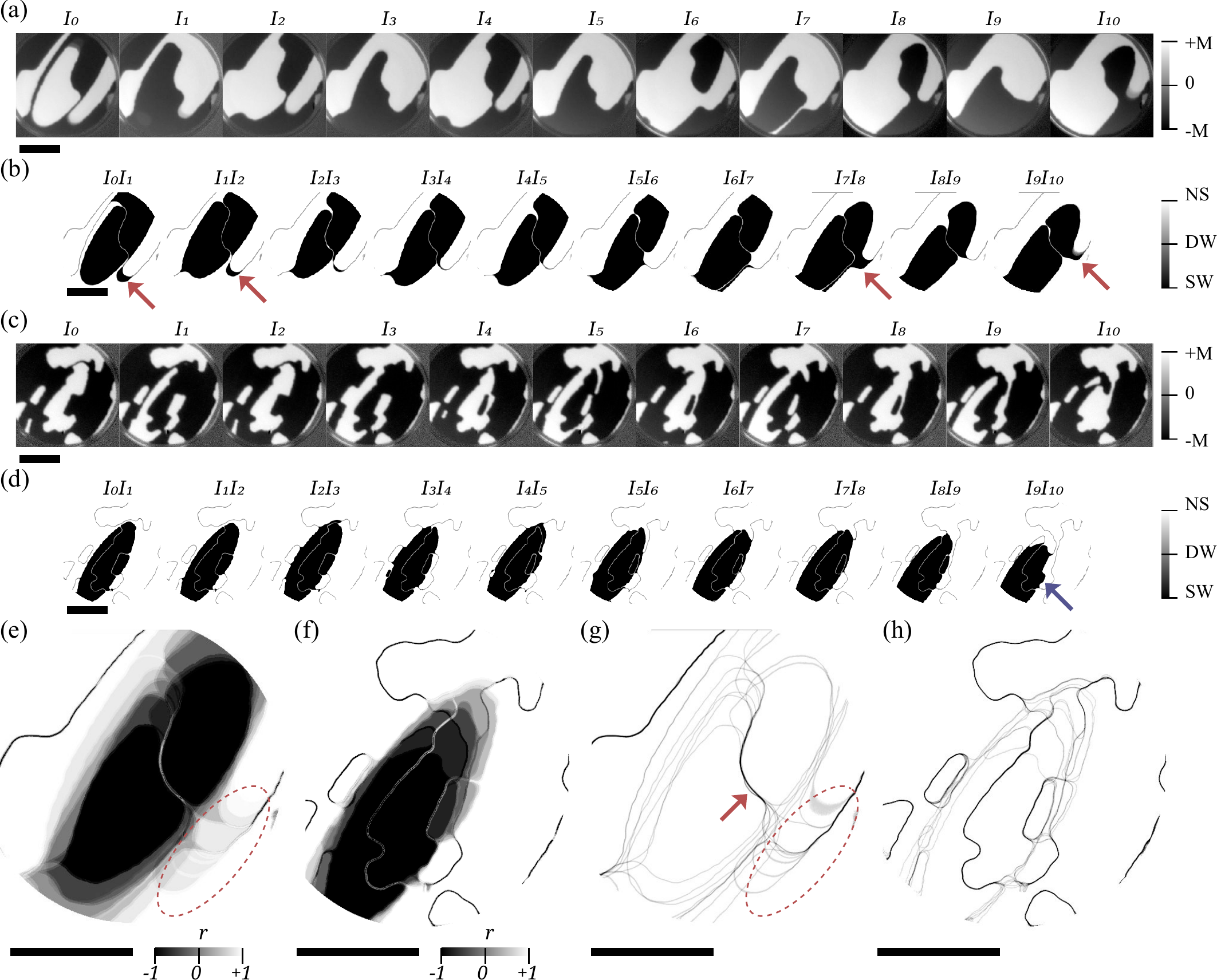}
\caption{\label{fig:SSTM}
(Color online)(a) Sequences of XMCD PEEM images $I_p$ taken after $p$
single laser pulse excitation above at T = 300~K and (c) below at T = 160~K
the magnetization compensation temperature \TM{} = 260~K of the Gd25FeCo
sample. The gray scale in the inset on the right indicates the
out-of-plane magnetization orientation.
(b) Sequences of image product $I_{p-1}I_p$ above and (d) below \TM{}.
The gray scale in the inset on the right indicates which gray level
corresponds to magnetization switching (SW), no switching (NS) or
domain wall (DW).(e) Correlation coefficient images $r$ derived from the sequences of 
single laser pulse excitation above and (f) below \TM{}.
(g) Average image $\langle I_p^2\rangle$ showing the domain wall positions
above and (h) below \TM{}.
Arrows and dashed ellipses indicate magnetization switching not
related to AOS and are discussed in the text. All scale bars are 20~$\mu$m.}
\end{figure}

In view of potential applications, the question of the repeatability of AOS is
essential. AOS was therefore studied on a multi-domain configuration were one
laser pulse excites several different magnetic domains at once. The magnetic
domain configuration before and after single linearly polarized laser pulse
exposure was recorded with static XMCD PEEM imaging.  Sequences of such $I_p$
images taken at the Fe $L_3$ edge for the Gd25FeCo sample at a temperature
above and below \TM{} in the absence
of any applied magnetic field are shown in Figs.~\ref{fig:SSTM} (a)
and (c) respectively. In those images, white (black) contrast corresponds to magnetic
domains whose out-of-plane magnetization has a positive (negative) projection on the X-ray
direction, as indicated by the gray scale in
Figs.~\ref{fig:SSTM} (a) and (c). In both cases, below and above
\TM{}, changes in the magnetic domains in the center of the images are seen.
To better emphasise the changes or the lack of them occurring in these
multi-domains configuration, the pixel by pixel product between two successive
images separated by a single linearly polarized laser pulse excitation
$I_{p-1}I_p$ is computed and shown in Fig.~\ref{fig:SSTM}(b) and (d).
Irrespective of the initial magnetic domain orientation, in the $I_{p-1}I_p$
image, a black contrast corresponds to a magnetization switching (SW),
a gray contrast to a domain wall (DW) and a white
contrast to an absence of changes, \textit{i.e.} no switching (NS), as
indicated by the gray scale in the inset. Visible in the product
of successive images $I_{p-1}I_p$ shown in Fig.~\ref{fig:SSTM}(b) and (d) is a
black elongated elliptical region at the center surrounded by a white region
unaffected by the laser pulses. This elongated elliptical shape corresponds to
the laser spot size seen at the 16\degree{} grazing incidence used in this
experiment. This black elongated region clearly corresponds to a laser induced
switching occurring equally for both magnetic domain orientations enclosed in
the laser spot size. Since this AOS seems to occur with every laser pulse, it
appears to be purely deterministic. 
To better quantify how deterministic this phenomenon of AOS really is, we
introduce the pixel-by-pixel Pearson product-moment
correlation coefficient $r$ for a sequence of XMCD images as:
\begin{equation*}
r = \frac{\sum_{p=1}^n I_{p-1}I_p}{\sqrt{\sum_{p=1}^n I_{p-1}^2}\sqrt{\sum_{p=1}^n I_{p}^2}},
\end{equation*}
where $I_p$ is the XMCD image after $p$ laser pulses in the sequence.
In the case of purely deterministic switching,
this correlation coefficient $r$ is -1, while in the absence of changes,
\textit{i.e.} no switching, $r$ = +1. In the event of an unrelated
domain configuration after every single laser pulse, such as in the case of
heating above \TC{}, $r$ = 0. Such correlation coefficient images $r$
calculated from the measured sequences are shown in Figs.~\ref{fig:SSTM}(e) and
(f) for a sample temperature above and below $T_\textrm{M}$ respectively.  The
darkest region in these images corresponds indeed to a correlation coefficient
of $r$ = -1, \textit{i.e.} a purely deterministic switching with each of the 10
laser pulse of the sequence, occurring both below and above \TM{}. It is also
evident that the spatial extent of this $r$ = -1 region is limited by
the spatial extent with which these 10 laser pulses overlap. Therefore, the pulse to
pulse pointing stability is the only extrinsic limitation to a somewhat
purely deterministic AOS.

However, there can also be intrinsic limitation such as domain walls, in particular at
the boundary
between the switching and non switching region of each laser pulse. For example, in
the case of the sample temperature above \TM{} shown in Fig.~\ref{fig:SSTM}(b),
the domain wall at the bottom of the laser pulse region is nearly continuously
moving in the same direction between successive images, as indicated by the red
arrows as well as the dashed ellipse in Fig.~\ref{fig:SSTM}(e). As this domain
wall is clearly outside the elongated elliptical region where AOS occurs, we
know that the laser fluence is too low to induce a deterministic AOS. In fact,
in the XMCD PEEM images $I_1$ and $I_{10}$ shown in Fig.~\ref{fig:SSTM}(a), one
can even see the domain wall motion occurring during the imaging which results
in an extended gray region rather than a either completely black or completely
white region.  This is indicative of a very low coercivity of the domain walls
at this temperature which favors thermally activated domain wall movements in the
otherwise non switching region and should be regarded as intrinsically limiting
the repeatability of the AOS. Comparing the domain size above and below \TM{},
as shown in Figs.~\ref{fig:SSTM}(a) and (c), one can immediately realize that
the coercivity is higher in the second case as the magnetic domains are
smaller, and thus more stable. Nevertheless, here some changes in the
domain configuration can also be seen at the edges of the AOS region, as indicated by
the blue arrow in the $I_9I_{10}$ image shown in Fig.~\ref{fig:SSTM}(d). The
small protuberance corresponds to a small black domain outside the AOS region
which disappeared between the images $I_9$ and $I_{10}$ shown in
Fig.~\ref{fig:SSTM}(c). This is likely the collapse of a too small domain
formed by the intersection of the existent domain pattern and the AOS region
created by the laser pulse. These processes of domain collapse and thermally activated
domain wall hopping should not be confused with AOS. In fact, they lower
the repeatability of AOS.

Inside the $r$ = -1 region, all magnetic domains are switching with every laser
pulse. However, it is unclear what is happening for the domain wall separating them
since the correlation coefficient $r$ is undefined there.
To visualize the various domain wall
position during the sequence of laser pulses, it
is best to look at the low intensity part of the average of the squared image
$\langle I_p^2\rangle$ shown in
Fig.~\ref{fig:SSTM}(g) and (h) for the sample temperature above and below \TM{},
respectively. In those $\langle I_p^2\rangle$ images, the darker the domain
wall, the less it moved during the sequence of laser pulses.  In the
case of the sample at a temperature above \TM{} shown in
Fig.~\ref{fig:SSTM}(g), some changes are visible at the domain wall inside the
switching region, as indicated by the red arrow. In the case below \TM{} shown
in Fig.~\ref{fig:SSTM}(h), no changes are visible, meaning that the domain wall
stayed in place within the 100~nm spatial resolution of the instrument.
Considering the low coercivity of this material, this is a rather surprising and
noteworthy feature of AOS.
Nevertheless, evidences for potential domain wall hopping well inside the $r$ =
-1 region are seen at least in one case, limiting the repeatability of the AOS.
Overall, apart from the difference in coercivity, very little differences are
seen between AOS below and above \TM{}.

\subsection{Time-resolved dynamics around \TM{}}

\begin{figure}[ht]
\includegraphics[width=\doublelargeur]{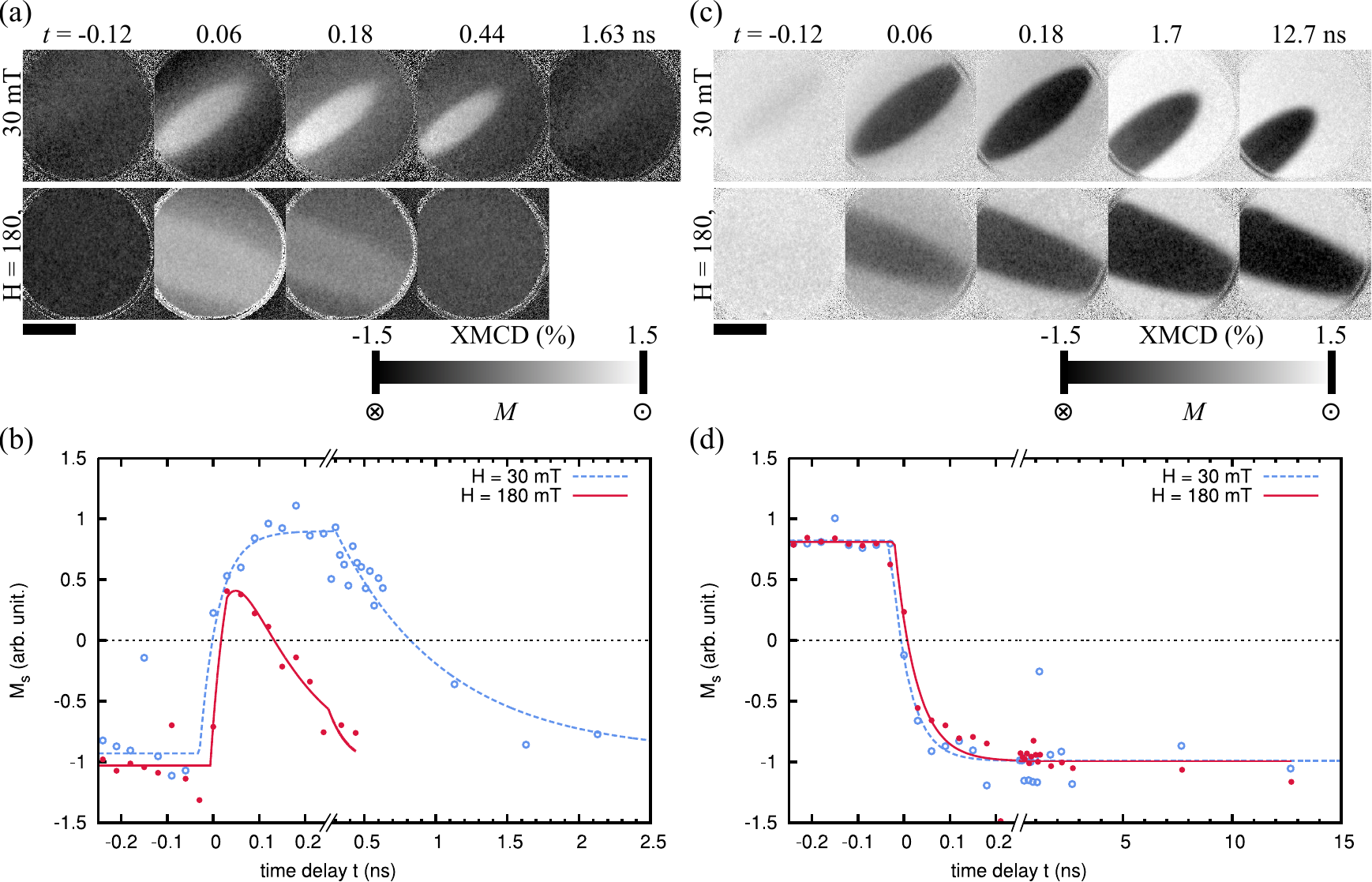}
\caption{\label{fig:TR_TM}
(Color online)(a) Time-resolved XMCD PEEM images on Gd25FeCo sample at different
time delays, for two different applied out-of-plane magnetic field of 30~mT and 180~mT,
measured at the Fe L$_3$ edge, at a temperature above \TM{} at T = 300~K and
(b) the extracted magnetization dynamics for each applied out-of-plane magnetic field.
(c) and (d) the same for a sample temperature below \TM{} at T = 160~K.
The scale bars are 20~$\mu$m.
}
\end{figure}

To gain more insight into the AOS and in particular into the role played by \TM{},
the magnetization dynamics in
this sample was investigated around \TM{}. For this, time-resolved XMCD PEEM measurements
were performed and the results are shown in Fig.~\ref{fig:TR_TM}, for a sample
temperature (a) above and (c) below \TM{}, and for a strong H = 180~mT and a
weak H = 30~mT out-of-plane magnetic field.  The magnetic field is used to
reset the sample magnetization to a well defined initial state, allowing for
stroboscopic measurement of the dynamics. The first thing to notice is that at
negative time delay $t$, i.e. before the laser pulse, the sample is saturated
for both applied magnetic fields, and that the orientation of the Fe sub-lattice
magnetization reverses between Fig.~\ref{fig:TR_TM}(a) and (c), meaning that the sample
is effectively on either side of the magnetization compensation temperature
\TM{} at the temperature used. From the time-resolved XMCD images, the
magnetization dynamics at the center can be extracted and is shown in
Figs.~\ref{fig:TR_TM}(b) and (d), for a sample temperature above and below \TM{},
respectively. In both cases, magnetization
switching occurs right after the laser pulse excites the sample.
Thus, within
the 70~ps time resolution of the experiments, no difference is seen in the
switching dynamics for either low or high magnetic field and either below or
above \TM{}. On the other hand, the relaxation towards the final state is
strongly influenced by both the applied magnetic field and the sample base
temperature. At a temperature above \TM{}, as shown in Fig.~\ref{fig:TR_TM}(b),
the reversed state is instable against the applied magnetic field, leading to a
fast relaxation towards the initial state, the faster the higher the field. It
is worth noting here that switching with a laser pulse against a field of
180~mT is thus possible, even though the relaxation is very fast, demonstrating
the impetuous by which this AOS occurs.~\cite{Ostler2012} Due to this fast
relaxation and the 70~ps long X-ray probe pulse length, a saturated switched
state is not observed. At temperatures below \TM{} as shown in
Fig.~\ref{fig:TR_TM}(d), the reversed state is now stable within the
illuminated area, indicating that the temperature in this region is now above
\TM{}. In this case, after the laser pulse, the applied magnetic field is now
stabilizing the reversed domain, leading to a very long life time.

\begin{figure}[ht]
\includegraphics[width=\largeur]{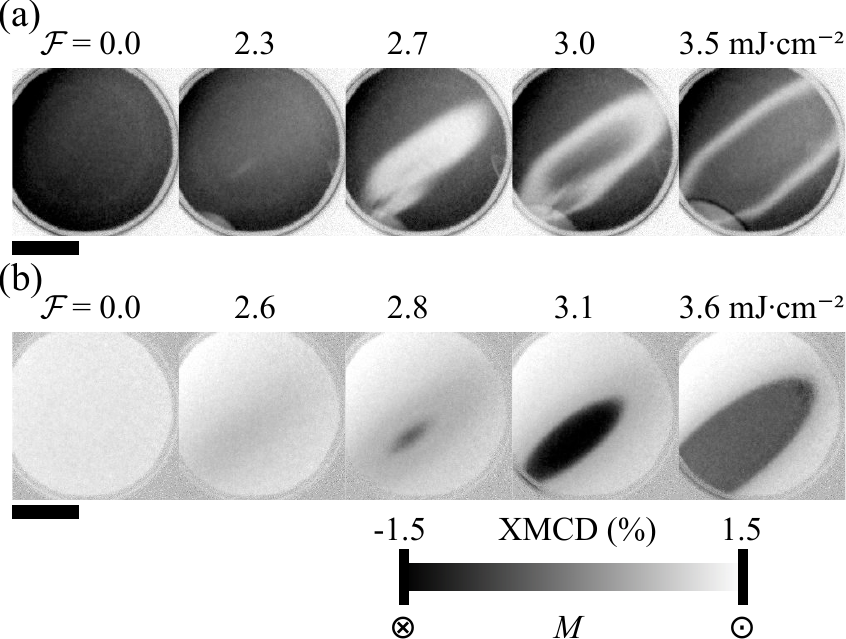}
\caption{\label{fig:fluTM}
(a) Time-resolved XMCD PEEM images taken at $t$ = 230~ps after the laser pulse
on Gd25FeCo sample above at T = 300~K and (b) below at T = 160~K the
magnetization compensation temperature \TM{} = 260~K, as a function of the
laser pump fluence. The static out-of-plane magnetic field was 30~mT.
The scale bars are 20~$\mu$m.}
\end{figure}

Time-resolved XMCD PEEM images taken at the same fixed time delay of $t$ =
+230~ps after the laser pulse on the same Gd25FeCo sample are shown in
Fig.~\ref{fig:fluTM}(a) above and (b) below \TM{}, as a function of the laser
pump fluence. A small static out-of-plane magnetic field of H = 30~mT was
applied to reset the sample after switching. This 30~mT magnetic field is small
enough to not hinder the AOS at this time scale as can been seen in
Fig.~\ref{fig:TR_TM}(b). While below \TM{}, the laser fluence can be increased
significantly without losing the AOS, the same is not true above \TM{}. There,
a small 10\% increase from 2.7 to 3.0 mJ$\cdot$cm$^{-2}$ is enough to bring the central
region of the laser spot into a demagnetized state. This effect is most
striking at the fluence of $\mathcal{F}$ = 3.5~mJ$\cdot$cm$^{-2}$ in
Fig.~\ref{fig:fluTM}(a), where the switched region forms a very thin 2~$\mu$m wide
ring around the laser pulse.  The AOS fluence switching window is thus reduced above
\TM{}, and this asymmetry of the switching window around \TM{} is consistent
with literature.~\cite{Vahaplar2009, Vahaplar2012, Medapalli2012} Part of this effect
might be attributed to the proximity with the Curie temperature $T_C$.

\subsection{Time-resolved dynamics far from \TM{}}

\begin{figure}[ht]
\includegraphics[width=\largeur]{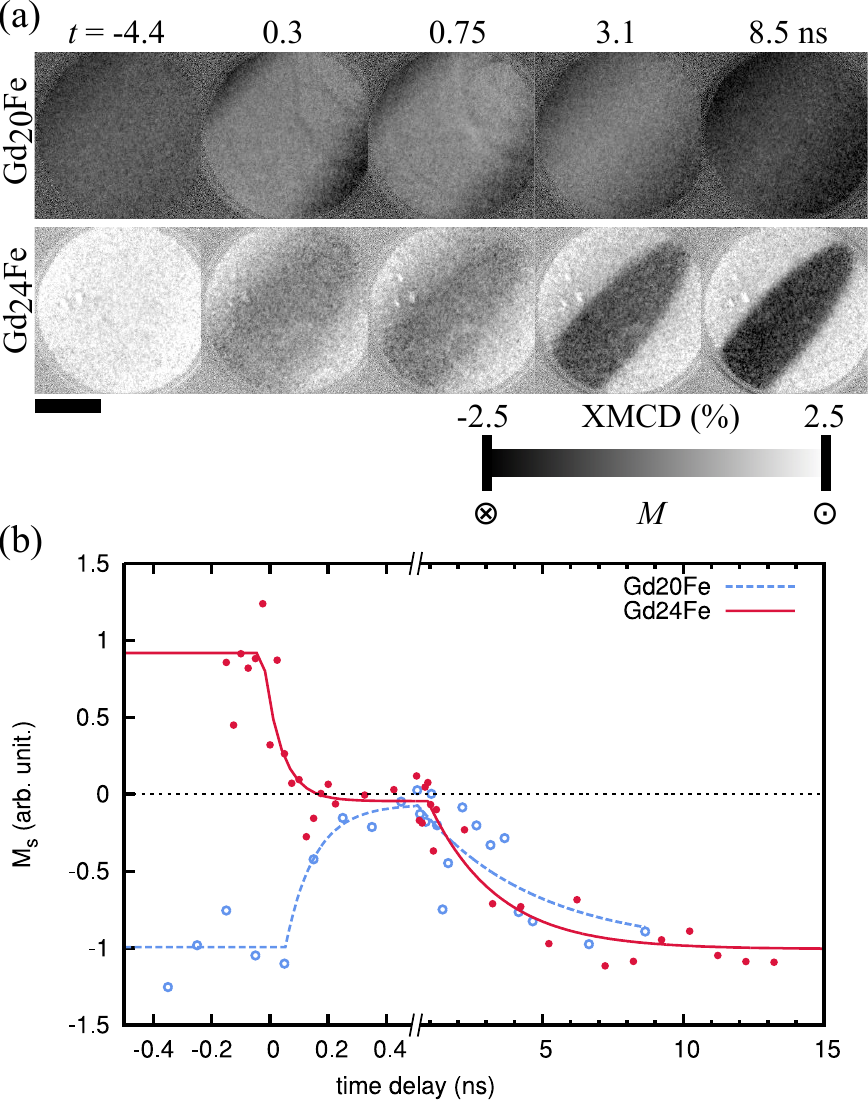}
\caption{\label{fig:GdFe}
(Color online)(a) Time-resolved XMCD PEEM images at various fixed
time delays and (b) extracted magnetization dynamics on Gd20Fe
(\TM{} around 0~K, $\mathcal{F}$ = 5.7~mJ$\cdot$cm$^{-2}$) and
Gd24Fe (\TM{} around 500~K, $\mathcal{F}$ = 3.9~mJ$\cdot$cm$^{-2}$) samples at
room temperature with a 30~mT of out-of-plane magnetic field.
The scale bar is 20~$\mu$m.}
\end{figure}

Due to the limited accessible temperature range in the PEEM, investigation of
the AOS far from \TM{} requires samples with different compositions. For this,
time-resolved XMCD PEEM measurement were thus performed on Gd20Fe with \TM{}
around 0~K and Gd24Fe with \TM{} around 500~K, under a small static
out-of-plane magnetic field of 30~mT. The results are shown in
Fig.~\ref{fig:GdFe}. For both samples, a time resolution limited
demagnetization process occurs. The samples then stay demagnetized for about
500~ps which is then followed by a slow dynamics on a time scale
of around 10~ns, towards
the initial state for Gd20Fe and towards the reversed state for Gd24Fe. This
reversal in the Gd24Fe sample shows that
there is an accessible magnetization compensation temperature \TM{}
in this sample below \TC{}, allowing the applied 30~mT out-of-plane
magnetic field to reverse the sample magnetization on a slow few nanoseconds
long time scale and eventually back to the initial state at even
longer time scale after cooling down.
For the Gd20Fe sample, the temperature is already above \TM{}
before the laser pulse, and therefore no magnetic field assisted switching
occurs. Looking at the XMCD PEEM images taken at fix time delay and shown in
Fig.~\ref{fig:GdFe}(a), it can be seen that in the case of the Gd20Fe sample,
the demagnetized region has a diffuse boundary, meaning that no magnetic domain
is actually formed. On the other hand, for Gd24Fe, at around 750~ps after the
laser pulse, a clear boundary appears in the heated region, which is seen in
Fig.~\ref{fig:GdFe}(a) at t = 3.1~ns. This very late formation of the reversed
domain in Gd24Fe and the absence of switching in Gd20Fe allow us to conclude
that no AOS window exists far from \TM{}.

\section{Discussion}

Determining if a system can display all-optical magnetization switching and to
which extent this AOS is deterministic are two questions of crucial importance,
for a better understanding of the phenomenon as well as in view of its potential
applications. In this context, sequences of XMCD PEEM images separated by
single linearly polarized laser pulse excitation on a multi-domain
configuration such as shown in Fig.\ref{fig:SSTM} can provide valuable
information. First of all, since linearly p-polarized laser pulses are equally absorbed
by each domain orientation, a direct comparison between what happens inside
each domain is possible.\cite{Khorsand2012} This is in contrast with multiple
circularly polarized laser pulses used in recent studies such as in
Refs.~\onlinecite{Mangin2014} and~\onlinecite{Lambert2014}, where such a
comparison can only be made after carefully taking into account the magnetic
circular dichroism of the material. Second, randomly demagnetized initial states are better
than saturated or artificially created domain states since no stray field is
created which could influence the switching. Third, the reversed domain
configuration in such case is known to be stable as well, therefore a collapse
of the reversed domain state because of too low coercivity or too high net
magnetization is not to be expected.\cite{Schubert2014} Finally, from such a
sequence of images, the actual reproducibility of AOS can be
measured using the Pearson product-moment correlation coefficient $r$ as
introduced.

From our analysis, it follows that the purely deterministic AOS
observed in the GdFeCo samples is limited by a number of extrinsic and intrinsic
effects. The largest limitation we observe in Fig.~\ref{fig:SSTM}(e) and (f) is
the pulse to pulse laser pointing stability which is extrinsic in nature to the
switching phenomenon itself. The second limitation observed is related to the
stability of the domain configuration. For example, at the edge of the laser
pulse, the overlap of the $r$ = -1 switching region with the preexistent domain
configuration can create domains which are too small to be stable, as seen in
Fig.~\ref{fig:SSTM}(d) $I_9I_{10}$. In addition, thermal activation of domain
walls can occur outside as well as inside the $r$ = -1 switching region, as seen
in Fig.~\ref{fig:SSTM}(g) and indicated by the arrow and dashed ellipse. Since
these two effects are related to the coercivity of the material, this
constitutes an intrinsic limitation to the repeatability of the AOS.
However, by understanding these limitations, we can envisage that engineered
materials can potentially alleviate these limitations. For example, in
patterned materials where each structure preferably host a single magnetic
domain, a purely deterministic switching would be maintained.

Regarding the role played by the magnetization compensation temperature \TM{}
on the AOS, we first of all confirm previous studies in that AOS occurs below
and above \TM{}.\cite{Vahaplar2012, Ostler2012} Single shot laser pulse
experiments shown in Fig.~\ref{fig:SSTM} as well as time-resolved measurements
of the magnetization dynamics shown in Fig.~\ref{fig:TR_TM} both reveal AOS
below as well as above \TM{}. However, there exists a clear difference between
switching below and above \TM{}, as shown in the fluence-dependent patterns
observed at $t$ = 230~ps in Fig.~\ref{fig:fluTM}.  In addition, for GdFe samples
with no or far from their \TM{}, no switching is observed, as shown in
Fig.~\ref{fig:GdFe}. This leads to the conclusion that while the existence of a
reachable \TM{} during the laser excitation is not a strict requirement to
observe AOS, sample compositions with \TM{} near room temperature are preferred. It
must be noted that in addition to \TM{}, an angular momentum compensation
temperature \TA{} also exists at a slightly higher temperature.\cite{Stanciu2006}
However, our experimental geometry with out-of-plane magnetic field does not
allow magnetization precession dynamics to be observed, precluding any
investigation of the effect of \TA{} on AOS. Finally, AOS is a very robust
switching mechanism as it can be realized against an opposing applied magnetic
field~\cite{Ostler2012}, as demonstrated experimentally here in the case of a
180~mT field in Fig.~\ref{fig:TR_TM}(b).

\section{\label{sec:conclusion}Conclusions}

In conclusion, using static and time-resolved PEEM microscopy with XMCD to
probe the sample magnetization upon laser excitation, important aspects of the
AOS have been revealed. Sequences of images after single linearly polarized
laser pulse excitation on a multi-domain configuration allow for the study of the
repeatability of the process by using the correlation coefficient as its
measure. It is found the AOS in the Gd25FeCo sample studied is nearly purely
deterministic. Moreover, intrinsic limitation from the low coercivity of the
material leading to thermally activated domain wall hopping could be alleviated
in patterned media. From the time-resolved measurement of the magnetization
dynamics, it is found that AOS occurs below and above \TM{}, while on the other
hand, no AOS occurs for sample temperatures far from it.  Strong reduction of
the fluence switching window occurs above \TM{} and is likely related to the
proximity with the Curie temperature \TC{}. Finally, AOS against an applied
magnetic field of 180~mT is demonstrated, illustrating the impetus by which AOS
occurs.

\begin{acknowledgments}
We thank the European Research Council under the European Union’s Seventh
Framework Programme FP7/2007--2013 (grants
NMP3-SL-2008-214469 (UltraMagnetron), FP7-NMP-2011-SMALL-
281043 (FEMTOSPIN) and 214810 (FANTOMAS)) for part of the
financial supports as well as the MEXT-Supported Program for the Strategic
Research Foundation at Private Universities, 2013--2017.
Part of this work was performed at the Swiss Light Source,
Paul Scherrer Institut, Villigen, Switzerland. We thank J. Honegger for his
support.
\end{acknowledgments}


%

\newpage{}

\end{document}